# Recent QCD results from LHC experiments

Celso Martínez Rivero
Instituto de Física de Cantabria (CSIC-UC)

On behalf of ATLAS and CMS collaborations.

The hard scatter from initial state hadrons to final state partons at LHC is reviewed with 2010 integrated luminosity. Plots showing certain quantities are shown as well as the powerful software tools for describing expectations.

## 1 Introduction

The Final states at hadron colliders are dominated by QCD processes. Indeed almost any new physics involves QCD, which is a major background for new physics searches. Thus, a better understanding of QCD gets an improved sensitivity to new physics.

Below we will see perturbative and non-perturbative QCD with some results on Jet properties, Inclusive Jet cross sections, Dijet cross sections, Multijet production, W/Z Jets and Isolated prompt photons in the first case and Underlying event and minium bias in the non-perturbative QCD. Finally a conclusions section will finish this paper.

## 2 Perturbative QCD

### 2.1 Jet properties

It is of high importance to study the MC description of Jets, which is important for QCD measurements. Both the charged particle multiplicity versus the jet $p_T$ (whose uncertainties from $p_T$ migration are associated with the jet energy scale) as well as the multiplicity distributions in intervals of $p_T$ and its comparisons with Pythia tunes. Figure 1 shows those kinds of distributions.

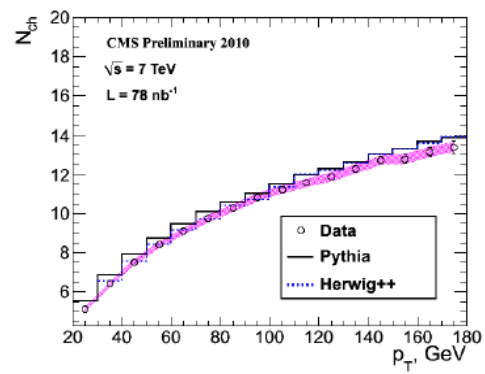

Fig.1: Mean charged multiplicity versus Jet $p_T$

A very active area for studies at high $p_T$ is the substructure of jets. There can be boosted decays of heavy particles (high mass jets with very near-symmetric sub-jets) or, on the other hand, QCD develops mass from gluon radiation (gives asymmetric sub-jets). Algorithms for measuring "fat" jets as applied to QCD jets consider large anti-kt jets (R=1) and the split of final two elements of the cluster.

The measured splitting scale for final cluster in QCD anti-kt jets is around 30 GeV as can be seen in figure 2.

CMS looked in a similar direction by using Algorithms for top tagging and W tagging with Jet pruning to search for Z' -> tt̄

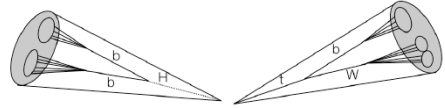

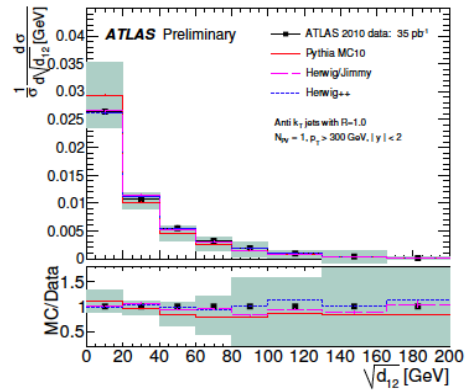

Fig:2 Final cluster splitting in QCD Jets.

### 2.2 Inclusive Jet Cross Section

The anti-$k_T$ jet algorithm is standard in ATLAS and CMS: soft particles first cluster with hard ones before clustering among themselves. Figure 3 shows $p_T$ up to 1 TeV and |y|<4.7. Cross sections vary by $10^{10}$ over all the $p_T$ range measured

Both experiments compare with pdfs baseline: CTEQ6.6 for ATLAS and PDF4LHC for CMS.

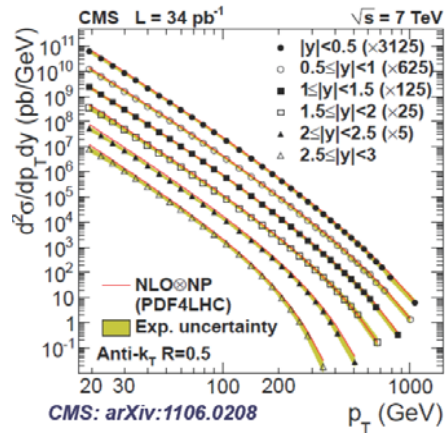

Fig:3 Cross section as function of $p_T$.

## 2.3 Dijet Cross Section

Dijet cross section has been calculated as function of $M_{jj}$ and $|y|_{max}$ up to 4 TeV in 5 intervals of $|y|$. Data and predictions are in good agreement with uncertainties of 10-15%. Figure 4 shows these distributions.

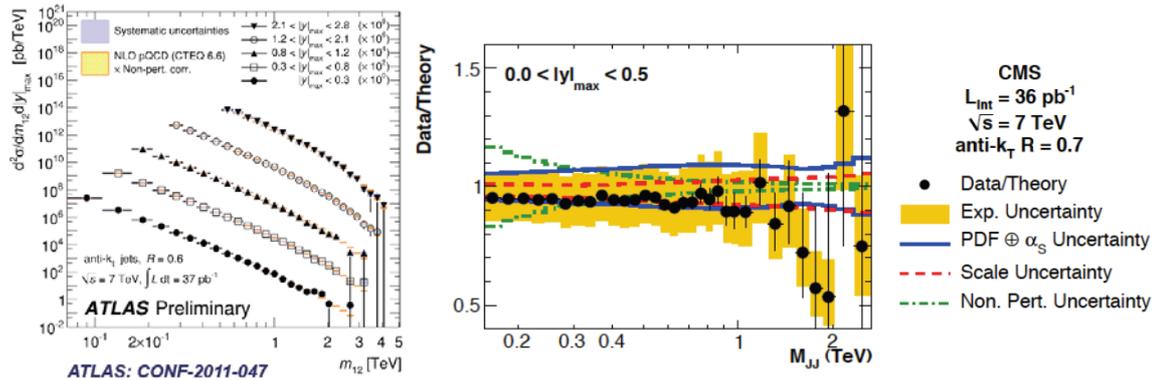

Figure 4: ATLAS dijet cross section as function of $|y|$ (left) and CMS ratio data/MC in the dijet cross section (right).

The Inclusive to exclusive dijet cross section ratio for jets with Pt>35 GeV and |y|<4.7 as function of their rapidity separation measured by CMS are also measured. CMS has proved that the inclusive cross section is 1.2-1.5 times larger than the exclusive one and that there are good agreement between data and Pythis6 predictions as seen in Figure 5.

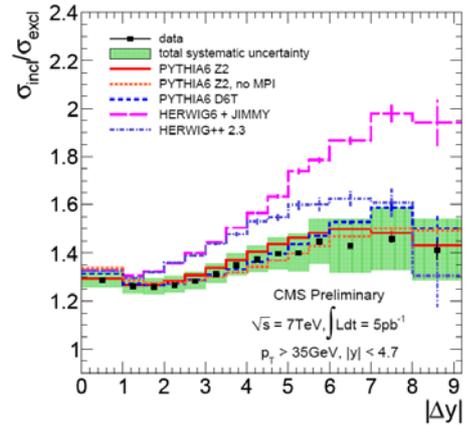

Figure5: Ratio of inclusive to exclusive cross section as function of |y|

## 2.4 Multijet Production

The same thing can be studied over multijets. If one defines $H_T$ as the sum of the $p_T$ of all the jets in the event, the ratio of 3-jet cross section versus 2-jet cross section over $H_T$ for $p_T$ >50GeV can be seen in Figure 6 (left) from CMS, while the right part compares the cross section as function of the number of jets for multiple MC. In this part, one can see as SHERPA provides de best normalization.

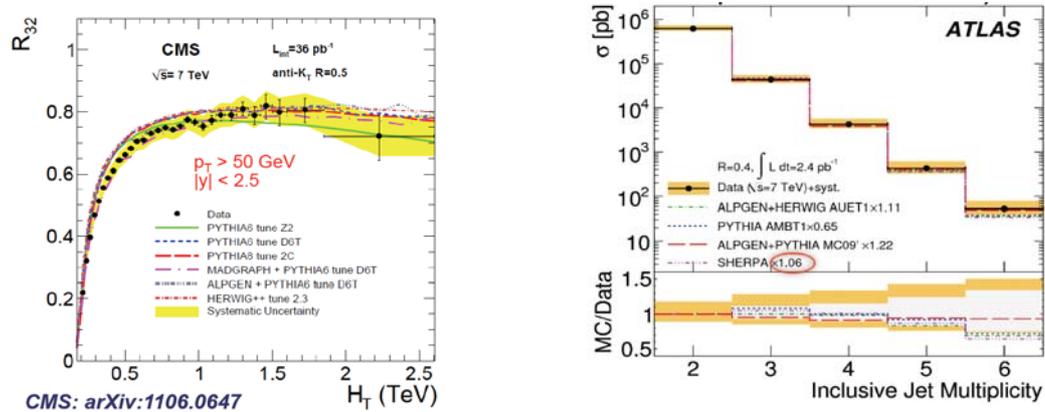

Figure 6: $R_{32}$ vs $H_T$ from CMS (left) and cross section as function of jet multiplicity from ATLAS (right)

## 2.5 W/Z + Jets

It is an important background for many searches (top, Higgs, SUSY). It does confront pQCD predictions as it gives a test of tree-level generators as Madgraph, ALPGEN and SHERPA.

There´s a good agreement with Madgraph and ALPGEN and also with NLO MCFM and BLACKHAT+SHERPA; however theer´s poor agreement with Pythia as shown in Figure 7.

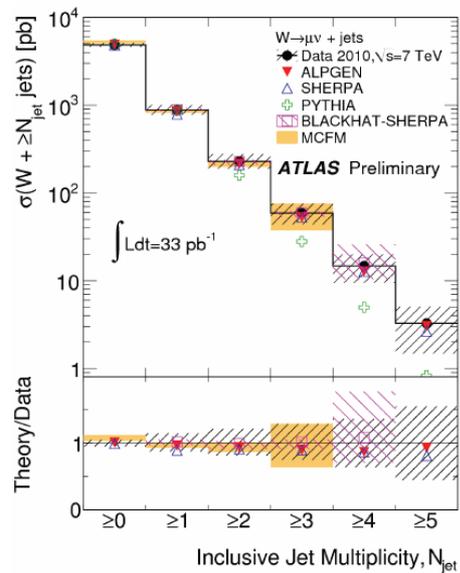

Figure 7: Cross section of W to n-jets as function of the number of jets.

## 2.5 Isolated prompt photons

It is a traditional channel for testing perturbative QCD**.** Photons can de direct or from parton fragmentation (one has to define an isolation cone around the photon to constraint second source). Both components are computed at NLO and implemented in JETPHOX MonteCarlo.

It is important to understand photon production at LHC, so it's a good case to measure photon detection and show that there is a good agreement between data/MC as is shown in figure 8. One sees there that experimental uncertainties are greater than the pdf while theory scale uncertainty is greater or of the order of the pdf.

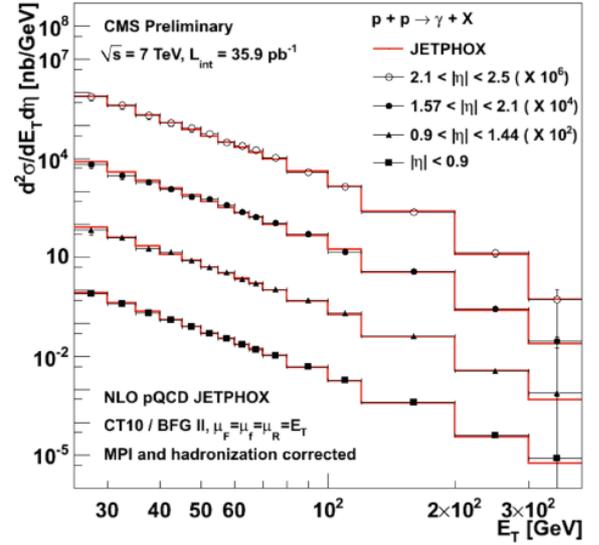

Figure 8: Photon cross section as function of $E_T$

# 3 Non-perturbative QCD

### 1.0 Underlying Event

Look at properties in 60 degrees azimuthal wedge transverse to leading track and plot the charge density as function of the leading tk and the <$p_T$> of charged tracks versus the p of the leading tk.

Results are in good agreement between experiments and with the underlying fit. One can see in Figure 9 a rapid with $p_T$, getting a plateau at 5 GeV

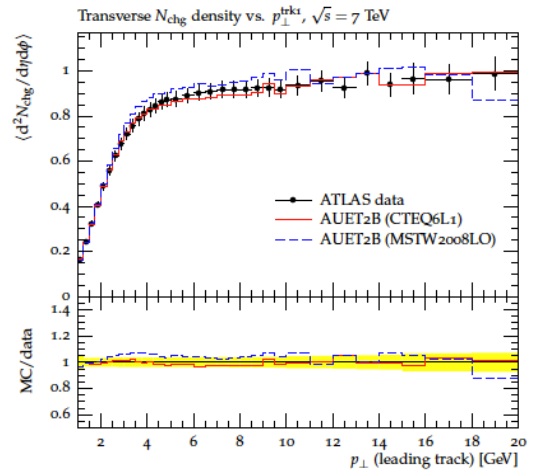

Figure 9 : Number of photons versus the $p_T$ of the leading track

## 4 Conclusions

We have presented a remarkable number of interesting events both from ATLAS and CMS, with common QCD software tools that unify the work and are essential for understanding data.

Most of the QCD predictions are in agreement with experiments, with 10-15% precision between data and MC. This precision tends to improve, as we know better detectors and will use larger event sample in the future.

## 5 References

- arXiv:1202.3548  QCD results from the LHC. Richard Nisius. New Trends in TERA Physics 2011
- arXiv:1106.3360 Studies of jet production with CMS in pp collisions at sqrt(s)=7 TeV. Panagiotis Katsas. $40^{th}$ symposium on multiparticle dymamics. University press Antwerp.